\documentstyle[prl,aps,psfig,multicol]{revtex}
\draft
\newcommand{\be}{\begin{equation}}
\newcommand{\ee}{\end{equation}}

\newcommand{\bary}{\begin{eqnarray}}
\newcommand{\eary}{\end{eqnarray}}
\begin{document}
\title{
Cosmic Rays Propagation in Bose Condensed Dark Matter}
\author{    
Sarira Sahu$^{*}$}
\address{ Physical Research Laboratory,
Navrangpura, Ahmedabad - 380 009, India}
\maketitle
\begin{abstract}
\noindent  We have calculated the dispersion relation for the high energy 
cosmic rays (protons, electrons and neutrinos) propagating in the Bose 
gas as well as Bose 
condensate medium of the pseudo-scalar
particles. For cosmic rays propagating in the Bose gas, the mass of the
particle will decrease and this is proportional to the boson density
in the medium. If the propagating fermion is massless then it 
will develop an imaginary mass and thus will be absorbed in the medium.
But if the condition $m^2/m_{\phi}p_0\simeq 1$ is satisfied, then the
cosmic rays will propagate freely without loosing energy in
the Bose condensate medium.\\
\noindent PACS number(s): 98.70.-f, 03.75.Fi, 95.35.+d
\end{abstract}

\begin{multicols}{2}
\section{Introduction}
Axions and majorons are the pseudo-scalar Goldstone bosons associated with
the Peccei-Quinn and the U(1)$_{B-L}$ symmetry respectively and they get mass
because of the spontaneous breaking of these symmetries\cite{ellis,valle}. 
Astrophysical and cosmological calculations allow the axion mass range 
$10^{-6}$ eV to $10^{-3}$ eV. Axions in this mass range
would have been produced non-thermally, as a Bose condensate
i.e., a classical coherent field oscillation of the axion
field\cite{preskil}. Because the axions are produced with small velocity, they
behave as cold dark matter (CDM), in spite of their small masses. Also the CDM
scenario for structure formation is at present the most promising with axions
and neutralinos being the leading CDM candidates. If axions provide the bulk
of the dark matter, they must comprise a significant fraction of the dark halo
of our own galaxy\cite{ipser}.
The existence of coherent axion mini clusters has been suggested\cite{kolb}. It
is shown that, axion clumps were formed in the universe, when the temperature
was about 1 GeV, because of the nonlinearity of the axion potential and
the inhomogeneity of coherent axion oscillation of the scale beyond the
horizon\cite{iwa}. 

It is expected that, in the present universe, there exist the axion
mini clusters and the axion boson stars as well as the incoherent 
axion gas as dark matter comprises these coherent axion clumps.
Majorons of KeV mass is also 
proposed as the candidate for the dark matter\cite{valle}.
As the majorons are also pseudo-scalar particle, they
can undergo Bose Einstein Condensation (BEC) and can form clusters in the
galactic halo. 

 The discovery of ultra-high energy cosmic rays with energy above
Greisen-Zatsepin-Kuzmin (GZK) cut-off ($\sim 5\times 10^{19}$ eV)\cite{gzk} are 
particularly mysterious since, if the cosmic rays are protons, then
photopion production caused by the resonant scattering process with the
microwave background should result in a rapid loss of energy.
It is believed that all the high energy cosmic rays are AGNs origin. The
propagation of these high energy particles in the inter galactic medium 
is also presently a puzzling phenomena.

In this present work we study the propagation of ultra-high energy cosmic
rays in the bosonic medium and BEC medium. The paper is organised as follows:
propagation of the fermion in the Bose gas is studied in section 2.
Section 3 deals with the propagation of fermion in the BEC medium. 
Ultra-high energy cosmic rays propagation, in Bose gas and in BEC medium is
discussed in section 4. In conclusions we  briefly summarize our results. 

\section{Fermion self-energy in  a Bose gas}

At finite temperature and density the physical processes take place in a heat
bath of particle and anti particles and the properties of the test particle
deviate from its vacuum values. The dispersion relation satisfied by the
free propagation of the particle is modified, in the presence of the heat
bath. The dispersion relation of the photon and leptons are well studied
in the finite temperature heat bath, particularly in astrophysical as well
as in cosmological scenario. However to my knowledge the particle propagation 
in Bose gas and in Bose condensed matter is not yet well studied. 

The fermion and pseudo-scalar interaction is given by the lagrangian density,
\be
{\cal L}_{int} = i g_p {\bar\psi}\gamma_5 \phi\psi
\ee
where $\phi$ is the pseudo-scalar field and $\psi$ is the fermion field.
The self energy for fermion in the presence of a heat bath is given by
\cite{jose},
\be
-i \Sigma = \int \frac{d^4k}{(2\pi)^4} 
\left (
i g_p\gamma_5 i S_F(p-k) i g_p \gamma_5 i D(k)
\right )
\label{self1}
\ee
where $S_F(p)$ and $D(k)$ are the fermion and the boson propagator. In real
time formalism of finite temperature field theory the fermionic and the
bosonic (scalar/pseudo-scalar) propagators are given by
\be
S_F(p) = (\not{p} +m) \left (\frac{1}{(p^2 -m^2)} + i \Gamma_F(p)\right )
\ee
and 
\be
D(k) = \frac{1}{k^2-m^2_{\phi}} - i\Gamma_B(k)
\ee
respectively. The quantities
\be
\Gamma_F(p) = 2 \pi \delta(p^2-m^2) n_F(p_0)
\ee
and
\be
\Gamma_B(k) = 2 \pi \delta(k^2-m^2_{\phi}) n_B(w).
\ee

Let us consider the propagation of a fermion in the medium. The real part of
the self-energy term gives the propagation of the particle in the medium, 
and the imaginary part gives the absorption. Also let us consider that,
only bosons (pseudo-scalars) are there in the medium. Then the real part 
of the self-energy term will be given by
\be
Re \Sigma = -g_p^2 \int \frac{d^4k}{(2\pi)^3}
({\not p} -{\not k} + m)
\frac{\delta (k^2-m_{\phi}^2) n_B(w)}{(p-k)^2 - m^2}.
\label{sf1}
\ee
The real part of the self energy can be expressed as
\be
-Re\Sigma = a{\not p} + b\not u
\ee
where $a$ and $b$ are two Lorentz-invariant functions and
$u^{\mu}=(1,0)$, $p^{\mu}=(p^0, -{\bf p})$
and $k^{\mu}=(w, -{\bf k})$.
Then the coefficients $a$ and $b$ are given by
\be
a=\frac{\Big (T_p - T_u p_0\Big )}{\bf p^2}
\label{aa}
\ee
and
\be
b=\left (\frac{p_0^2}{\bf p^2} -1 \right) T_u - \frac{p_0}{\bf p^2} T_p,
\label{bb}
\ee
where $T_p$ and $T_u$ are given as
\be
T_p=-\frac{1}{4} Tr(\not{p} Re\Sigma)
\label{tp}
\ee
and
\be
T_u=-\frac{1}{4} Tr(\not{u} Re\Sigma).
\label{tu}
\ee
In general the dispersion relation for the fermion is given by
\be
det(\not{p} - m - \Sigma) = 0
\ee
and this gives rise to
\be
(p_0 (1+a) + b)^2 =m^2 + 2 (1+a)^2 {\bf p}^2.
\label{disp}
\ee
By substituting eq.(\ref{sf1}) in eqs.(\ref{tp}) and 
(\ref{tu}) and the simplifying them we obtain 
\bary
T_p &=&
-\frac{g_p^2}{32\pi^2 |{\bf p}|}\int \frac{k dk}{w_k}
{n_B(w_k)} \left [
(p_0^2-{\bf p}^2 +m^2-m_{\phi}^2) \right.
\nonumber\\
&& \left.\times
\ln {\left |\frac{D_2 + \sigma}{D_2-\sigma} \right |}
+ 2 \sigma \right ]
\eary
and
\be
T_u = -\frac{g_p^2}{(2\pi)^2}\int \frac{k dk}{2 w_k} (p_0-w_k)
n_B(w_k) \ln \left |\frac{D_2 + \sigma}
{D_2-\sigma} \right |,
\ee
where $w_k=\sqrt{k^2+m_{\phi}^2}$, 
\be
D_2=p_0^2-{\bf p}^2 -m^2-m_{\phi}^2 - 2 p_0 w_k
\ee
and $\sigma= 2 |{\bf p}||{\bf k}|$.
For fermions  having $p_0$ and $|{\bf p}|$ $>> ~m,~ m_{\phi}$ and
$|{\bf k}|$ the lograthmic term will vanish $i.e.$
\be
\ln \left |\frac{D_2 + \sigma}
{D_2-\sigma} \right |\simeq 0.
\ee
This gives
\be
T_p=
-\frac{g_p^2}{8\pi^2}\int \frac{k^2 dk}{w_k} n_B(w_k)
\label{int}
\ee
and $T_u\simeq 0$. Putting $T_p$ and $T_u$ in eqs.(\ref{tp}) and (\ref{tu})
we obtain 
\be
b=-a p_0.
\label{boseb}
\ee 
For $exp{(\beta\sqrt{k^2+m^2_{\phi}})} >> 1$ we can write the integrand
in eq.(\ref{int}) as a sum. Then
\bary
\int \frac{k^2 dk}{w_k} n_B(w_k)
&=&\frac{1}{\beta^2}\sum_{n=0}^{\infty} \int_{\alpha}^{\infty} 
\sqrt{x^2 - \alpha^2} e^{-(n+1) x} dx\nonumber\\
&=& \frac{\alpha}{\beta^2}
\sum_{n=0}^{\infty} \frac{K_{-1}\left (\alpha(n+1)\right )}{n+1},
\label{tp1}
\eary
where $\alpha=m_{\phi}\beta$.
In the limit $\alpha(n+1)\rightarrow\infty$ the modified Bessel function
\be
K_{-1}(\alpha(n+1))\simeq \sqrt{\frac{\pi}{2 \alpha (n+1)}} e^{-(n+1) \alpha}
\left ( 1+ \frac{3}{8\alpha}\right )
\ee 
Putting this in eq.(\ref{tp1}), we get
\be
\int \frac{k^2 dk}{w_k} n_B(w_k)
=\sum_{n=0}^{\infty}\sqrt{\frac{\pi \alpha}{2\beta^4 (n+1)^3}} 
e^{-(n+1) \alpha}
\left ( 1+ \frac{3}{8\alpha}\right ).
\ee
Keeping the leading order
term in $n$ and for $m_{\phi}\beta >> 1$ we have
\be
T_p = 
-\frac{g_p^2}{4 m_{\phi}} n_0,
\ee
where
\be
n_0=
 \Big (\frac{m_{\phi}}{2\pi\beta}\Big )^{3/2}
e^{-m_{\phi}\beta}
\ee
is the number density of bosons in the medium.
Then the dispersion relation 
in eq.(\ref{disp}) is simplified to
\be
p_0^2-{\bf p}^2 - m^2 = {\bf p}^2 (a^2 + 2 a).
\ee
Keeping terms up to order $a$ in the above equation 
the dispersion relation is
\be
p_0^2-{\bf p}^2=m^2 - \frac{g_p^2}{2 m_{\phi}} n_0.
\label{bosegas}
\ee
Thus in a bosonic medium the propagating particle mass square decreases 
by an amount $\delta m^2={g_p^2}n_0/{2 m_{\phi}}$.
If the propagating fermion is massless then it is Landau damped, because
it will acquire an imaginary mass. On the other hand if the condition
$m^2={g_p^2}n_0/{2 m_{\phi}}$ is satisfied then the fermion will propagate
like a massless particle in the vacuum.

\section{Fermion self-energy in BEC medium}

For BEC of the pseudo-scalar bosons, the
number density $n_0$ can be expressed as\cite{landau}
\be
n_B(k_0) = (2\pi)^3 n_0 \delta^3({\bf k})
\label{bec}
\ee
where $n_0$ is the particle density in the condensate medium, with zero
momentum.
Then putting eq.(\ref{bec}) in  eq.(\ref{sf1}) we get
\be
T_p = - g_p^2 \frac{(p_0^2 -{\bf p}^2 - p_0 m_{\phi})}
{(p_0^2 - {\bf p}^2 + m_{\phi}^2 - 2 p_0 m_{\phi} - m^2)} \frac{n_0}
{2 m_{\phi}}
\ee
and 
\be
T_u= - g_p^2 \frac{(p_0 - w)}
{(p_0^2 - {\bf p}^2 + m_{\phi}^2 - 2 p_0 m_{\phi} - m^2)} \frac{n_0}
{2 m_{\phi}}.
\ee
Then putting the values of $T_p$ and $T_u$ in 
eq.(\ref{aa}) and (\ref{bb})  we obtain for the BEC medium 
\be
a=\frac{C_p}{(p_0 -m_{\phi})^2 -{\bf p}^2 - m^2}
\ee
and 
\be
b=-a m_{\phi},
\ee
where $C_p={g_p^2 n_0}/{2 m_{\phi}}$.
For BEC medium $b$ is proportional to the pseudo-scalar mass whereas for
Bose gas it is proportional to the energy of the propagating fermion as
shown in eq. (\ref{boseb}).
Putting the value of $b$ we obtain
\be
\left (p_0 (1 +a) - a m_{\phi}\right)^2 = m^2 + (1 + a)^2 {\bf p}^2.
\ee
Keeping terms up to order $a$ the dispersion relation will be
\be
p_0^2 - {\bf p}^2 =m^2 +
\frac{2 C_p m_{\phi} p_0 - 2 m^2 C_p}{(p_0^2 -{\bf p}^2 - m^2) + m_{\phi}^2 -
2 p_0 m_{\phi}}.
\ee
Writing $p_0^2 -{\bf p}^2 - m^2 = \delta m^2$
we get
\be
\delta m^2 = \frac{2 C_p (m_{\phi}p_0 - m^2)} {\delta m^2 +
(m_{\phi}^2 - 2 p_0 m_{\phi})}.
\ee
It is a quadratic equation in $\delta m^2$, so $\delta m^2$ has
two solutions. But the physical solution is the one, which goes to zero
for $ C_p$ goes to zero. So the value of $\delta m^2$ is given by
\be
\delta m^2 = -\frac{2 C_p \left (p_0 - \frac{m^2}{m_{\phi}}\right )}
{2 p_0 - m_{\phi}}.
\ee
For $p_0 >> m_{\phi}$ the above relation will be
\be
\delta m^2 = -\frac{C_p}{p_0} \left (p_0 - \frac{m^2}{m_{\phi}}
-\frac{m^2}{2 p_0} +\frac{m_{\phi}}{2}
\right )
\ee
Now the dispersion relation for the fermion propagating in the BEC 
medium is 
\be
p_0^2-{\bf p}^2 = 
m^2 - \frac{g_p^2 n_0}{2 m_{\phi}} \left (1 - \frac{m^2}{m_{\phi} p_0}
-\frac{m^2}{2 p^2_0} +\frac{m_{\phi}}{2p_0}
\right )
\label{fdis}
\ee
where is $n_0$ is the number density of the pseudo-scalar particles in the
medium. 
Comparison of eq.(\ref{bosegas}) with eq.(\ref{fdis}) shows that, 
we do not have a pure BEC medium, because the
second term in the right hand side of eq.(\ref{fdis}) $i.e.$ 
$g_p^2 n_0/2 m_{\phi}$ comes from the 
Bose gas contribution. Again 
the eq.(\ref{fdis}) clearly shows that, for a massless fermion 
propagating in the BEC medium, the particle will develop a negative mass
square value of magnitude $g_p^2 n_0(1+m_{\phi}/2 p_0)/2 m_{\phi}$ and will
be absorbed in the medium.
Now let us compare the contributions of different terms (within the
bracket of eq.(\ref{fdis})). For high energy particle $p_0 >> m_{\phi}$. So
the term $m_{\phi}/2 p_0$ almost does not contribute. 
For particle with $p_0 >> m$
the term $m^2/2 p_0$ is also small. So only the first and the second terms 
with in the bracket are contributing. Thus we have
\be
p_0^2-{\bf p}^2 \simeq
m^2 - \frac{g_p^2 n_0}{2 m_{\phi}} \left (1 - \frac{m^2}{m_{\phi} p_0}
\right ).
\ee

Let us consider different situations, which might arise depending on the
mass and energy of the propagating particle.
For massive fermions
if $m^2/m_{\phi} >> p_0$ condition is satisfied 
then the particle mass will be increased
by an amount $g_p^2 n_0 m^2/2 m^2_{\phi} p_0$ and 
the BEC contribution will dominate over the Bose gas one. On the contrary
if $m^2/m_{\phi} << p_0$ then, the particle mass will be reduced by an amount
$g_p^2 n_0/2 m_{\phi}$ which is purely due to the Bose gas nature of 
the medium and BEC contribution is negligible. For $p_0\simeq m^2/m_{\phi}$, 
the particle will propagate like a free particle in the BEC medium 
without loosing any energy.

\section{Cosmic Rays propagation}

Active galactic nuclei are found  to be the origin of the highest 
energy cosmic rays, like photons, protons, electrons and neutrinos\cite{thomas}.
The recent discovery of cosmic rays events above Greisen-Zatsepin-Kuzmin
(GZK) cut-off ($\sim 5\times 10^{19}~eV$)\cite{gzk} by the AGASA\cite{agasa}, 
Fly's Eye\cite{flyeye}, 
Haverah Park\cite{hp} and Yakutsk collaborations\cite{yak} 
is an outstanding puzzle in cosmic
rays physics. If these cosmic rays are protons with energy more than GZK
cut-off then, interaction with the
cosmic microwave background radiation (CMBR), would cause rapid loss of
energy by photopion production and consequently depletion of the observed
flux of these particles\cite{gzk}. 
For every mean free path $\sim 6$ Mpc of travel, 
the proton loses 20\% of its energy on average\cite{halzen}. 
Since AGNs are hundreds of megaparsecs away, the energy requirement for
a proton which arrives at earth with a super GZK energy is extremely high.
Presently we do not have a reliable theoretical model to explain the 
origin and propagation of such high energy cosmic rays in the galactic medium.
On the other hand ultra-high energy neutrinos will escape first and bring 
us first-hand information regarding the source. Also these AGN neutrinos 
can be used to study the properties of neutrinos themselves\cite{thomas}.

Here we will see how the propagation of these ultra-high energy cosmic rays
(protons, electrons and neutrinos) are affected by 
axion and majoron dark matter in the galactic halos.
 
\subsection{Bose gas}
  Let us consider ultra-high energy cosmic rays proton propagation 
in the axion gas. Then for proton-axion coupling $g_p\sim 10^{-10}$, 
$m_{\phi}=m_a\simeq 10^{-3} ~eV$\cite{ellis} and axion number density 
$n_0\simeq 3\times 10^{13}/cm^3$
we obtain 
$g_p^2 n_0/2 m_a\simeq 1.2\times 10^{-18}~~eV^2$ and this is much smaller
than proton mass square $m_p^2$. Also this term is much smaller than
electron mass square $m_e^2$ (we assume the electron-axion coupling same
as the proton-axion coupling).
So the axion gas has  no effect
on the propagation of the cosmic rays protons and/or electrons.
Secondly let us consider the neutrino propagation in the majoron gas, then
taking neutrino-majoron coupling 
$g_p\sim 10^{-4}$, $m_J\simeq 1$ KeV\cite{dolgov} and the majoron 
density same as the axion density, we obtain
$g_p^2 n_0/2 m_J\simeq 1.2\times 10^{-12}~~eV^2$. For non-zero neutrino mass
the majoron gas has also no effect on the neutrino propagation.

\subsection{BEC medium}

 For proton  propagating in the axion Bose condensate, we have
$m_p^2/m_a\simeq 10^{21}~ eV$ which is above the GZK cut-off. 
For proton energy $p_0 << 10^{21}~eV$ 
the proton mass will increase by an amount
$g_p^2 n_0 m_p^2/(2 p_0 m_a^2)$. 
Similarly for electron 
propagating in the BEC of axion, we have
$m_e^2/m_{\phi}\simeq .25\times 10^{15}~eV$. In this case also for 
electron energy $p_0 < 0.25\times 10^{15}~eV$ the electron mass
will increase in the medium and for $p_0 >> 0.25\times 10^{15}~eV$ the mass
will decrease. If $m_p^2/{m_a p_0}\simeq 1$, then
the Bose gas contribution will be cancelled by the BEC contribution and
the proton will propagate like a free particle in the vacuum, without
loosing its energy. 

For neutrino propagating in the majoron condensate medium, we have 
$m^2_{\nu}/m_J\simeq 10^{-1}~eV$, for taking neutrino mass $m_{\nu}=10 ~eV$
and $m_J=10^3~eV$. The cosmic ray neutrinos have energy as high as
$10^{21}~eV$. So $m_{\nu}^2/{m_J p_0}\simeq 10^{-22}$; hence 
only the majoron Bose gas will contribute for the neutrino mass and the
neutrino mass term in eq.(\ref{fdis}) will be
$(m^2-g_p^2 n_0/2 m_J)$. This reduction in neutrino mass is very small to
account for.

\section{Conclusions}

    We have calculated the dispersion relation for the fermion 
propagating in the Bose gas and the BEC of the pseudo-scalar particles.
It shows that, if the fermion is massless then, it will develop an imaginary
mass both in the Bose gas and in BEC medium, 
which is proportional to the number 
density of the bosons (Bose condensate) in the medium. For cosmic rays  
of energy below $m^2/m_{\phi}$, the mass of the fermion will increase,
on the other hand for energy above $m^2/m_{\phi}$ the mass will decrease.
Interesting situation arises, when $m^2/m_{\phi}p_0\simeq 1$. In this
situation the Bose gas contribution will be cancelled by the BEC contribution
and the particle will propagate freely without loosing its energy. But we
found that the contribution of the axion gas or BEC of axion to proton mass
is very small. Similarly majoron has also very small contribution to the
neutrino mass. This small contributions are solely due to the
low number density of these particles in the galactic halo.

It is a great pleasure to thank Dr. S. Mohanty for many
useful discussions.

\end{multicols}
\end{document}